\documentclass[%
 reprint,
 amsmath,amssymb,
 aps,
 prl
]{revtex4-1}

\usepackage{graphicx}
\usepackage{dcolumn}
\usepackage{bm}
\usepackage{amsmath,amssymb,mathrsfs}

\usepackage{amsfonts,bm}
\usepackage{amsmath}
\usepackage{amssymb}
\usepackage{amsthm}
\usepackage{graphicx}
\usepackage{graphics}
\usepackage{subfigure}
\usepackage{color}
\usepackage{bm}
\usepackage{hyperref}
\usepackage{rotating}
\usepackage{comment}

\usepackage[colorinlistoftodos]{todonotes}
\usepackage{esint}

\newcommand{\be}{\begin{equation} }
\newcommand{\ee}{\end{equation} }
\newcommand{\ba}{\begin{eqnarray} }
\newcommand{\ea}{\end{eqnarray} }

\newcommand{\bpm}{\begin{pmatrix}}
\newcommand{\epm}{\end{pmatrix}}
\newcommand{\bmm}{\begin{matrix}}
\newcommand{\emm}{\end{matrix}}

\newcommand{\la}{\label}
\newcommand{\p}{\partial}
\newcommand{\bea}{\begin{eqnarray}}
\newcommand{\eea}{\end{eqnarray}}

\def\blue{\color{blue}}

\def\Xint#1{\mathchoice
   {\XXint\displaystyle\textstyle{#1}}%
   {\XXint\textstyle\scriptstyle{#1}}%
   {\XXint\scriptstyle\scriptscriptstyle{#1}}%
   {\XXint\scriptscriptstyle\scriptscriptstyle{#1}}%
   \!\int}
\def\XXint#1#2#3{{\setbox0=\hbox{$#1{#2#3}{\int}$}
     \vcenter{\hbox{$#2#3$}}\kern-.5\wd0}}

\def\dashint{\Xint-}

\makeatother

\begin{document}


\title{Free surface variational principle for an incompressible fluid with odd viscosity}
 \author{Alexander G.~Abanov}
\affiliation{Simons Center for Geometry and Physics and Department of Physics and Astronomy, Stony Brook University, Stony Brook, NY 11794, USA}
 \author{Gustavo M. Monteiro}
 \affiliation{Instituto de F\'isica Gleb Wataghin, Universidade Estadual de Campinas-UNICAMP, 13083-859 Campinas, SP, Brazil}

\begin{abstract}
We present variational and Hamiltonian formulations of incompressible fluid dynamics with free surface and nonvanishing odd viscosity. We show that within the variational principle the odd viscosity contribution corresponds to geometric boundary terms. These boundary terms modify Zakharov's Poisson brackets and lead to a new type of boundary dynamics. The modified boundary conditions have a natural geometric interpretation describing an additional pressure at the free surface proportional to the angular velocity of the surface itself. These boundary conditions are believed to be universal since the proposed hydrodynamic action is fully determined by the symmetries of the system. 
\end{abstract}

\pacs{Valid PACS appear here}
\maketitle


\paragraph{\textbf{Introduction}.} 
Variational principle in hydrodynamics have a long history. We refer to Ref.~\cite{mobbs1982variational} and references therein for an introduction to the topic. In particular, the Luke's variational principle (LVP) is a variational principle of an inviscid and incompressible fluid with a free surface \cite{friedrichs1934minimumproblem,luke1967variational}. LVP provides both bulk hydrodynamic equations for an irrotational flow as well as kinematic and dynamic boundary conditions at the free surface boundary \cite{luke1967variational}. Such principle was later extended to include surface tension and bulk vorticity (for a recent summary see \cite{kolev2006variational}). In this letter, we present a further extension of LVP which accounts for the presence of odd viscosity in isotropic two-dimensional fluids with broken parity. 

In three dimensions, parity odd terms in the viscosity tensor were known for a long time in the context of plasma in a magnetic field~\cite{landau1987fluid} and in hydrodynamic theories of superfluid He-3A \cite{helium-book}, where the fluid anisotropy plays a major role. In two dimensions however the odd viscosity is compatible with isotropy of the fluid \cite{avron1998odd}. The odd viscosity is the parity violating non-dissipative part of the stress-strain rate response of a two-dimensional fluid. The recent interest in odd viscosity is motivated by the seminal paper by Avron, Seiler, and Zograf \cite{avron1995viscosity} where it was shown that, in general, quantum Hall states have non-vanishing odd viscosity. The role of odd viscosity (a.k.a. Hall viscosity) in the context of quantum Hall effect has been an active area of research~\cite{tokatly2006magnetoelasticity,tokatly2007new,tokatly2009erratum, read2009non,haldane2011geometrical,haldane2011self,hoyos2012hall, bradlyn2012kubo, yang2012band,abanov2013effective,hughes2013torsional, hoyos2014hall, laskin2015collective, can2014fractional,can2015geometry,klevtsov2015geometric,klevtsov2015quantum, gromov2014density, gromov2015framing, gromov2016boundary, scaffidi2017hydrodynamic, andrey2017transport,alekseev2016negative,pellegrino2017nonlocal}, but is out of the scope of this work.

In the Ref.~\cite{avron1998odd}, Avron has initiated the search for odd viscosity effects in classical 2D hydrodynamics. These effects are subtle in the case when the classical two-dimensional fluid is incompressible. Recent works have outlined some of observable consequences of the odd viscosity for incompressible flows~\cite{wiegmann2014anomalous,lapa2014swimming, banerjee2017odd, ganeshan2017odd,lucas2014phenomenology,abanov2018odd}. In particular, in Ref.~\cite{abanov2018odd} the equations governing the Hamiltonian dynamics of surface waves were derived in the case where bulk vorticity is absent.

Let us start by summarizing the main equations of an incompressible fluid dynamics with odd viscosity. In the following we assume that the fluid density is constant and take it as unity. We also neglect all thermal effects. Then, the hydrodynamic equations are the incompressibility condition and the Euler equation
\begin{eqnarray}
	\bm{\nabla}\cdot\bm{v} &=& 0\,,
 \label{eq:incompressibility} \\
	\p_t\bm{v}+ (\bm{v}\cdot {\bm \nabla})\bm{v} &=&  \bm{\nabla}\otimes \bm{T}\,.
 \label{eq:Euler}
\end{eqnarray}
Here, $\bm{v}(\bm x,t)$ is a two-component velocity vector field and $\bm{T}$ is the stress tensor of the fluid. In components the r.h.s. of the Euler equation (\ref{eq:Euler}) reads $(\bm{\nabla}\otimes \bm{T})_i=\nabla_j T_{ij}$. In flat space and in Cartesian coordinates, the stress tensor assumes the following form
\begin{eqnarray}
	T_{ij} =-\delta_{ij}p  +\nu_o(\p_iv_j^*+\p_i^*v_j)\,.
 \label{eq:Tij}
\end{eqnarray}
The first term of (\ref{eq:Tij}) is standard and describes the contribution to the stress from isotropic fluid pressure $p$. The second term, however, is quite different from the conventional dissipative shear viscosity $\nu_e(\p_iv_j+\p_jv_i)$ (here $\nu_e$ is shear or ``even'' viscosity coefficient). The last term  of (\ref{eq:Tij}), instead, is the contribution of the odd viscosity, with $\nu_o$ being the kinematic odd viscosity coefficient. Differently from $\nu_e$, we can assign either sign to the odd viscosity $\nu_o$, since it multiplies a dissipationless term. In (\ref{eq:Tij}) and in the following we use the ``star operation'' so that the vector $\bm{a}^*$ is the vector $\bm{a}$ rotated $90^{\circ}$ clockwise or in components $a_i^*\equiv \epsilon_{ij}a_j$. This operation explicitly breaks parity and a non-vanishing $\nu_o$ is only allowed in parity breaking fluids.

Euler equation (\ref{eq:Euler}) with the stress tensor (\ref{eq:Tij}) takes the form of the Navier-Stokes equation with odd viscosity term replacing the conventional viscosity term
\begin{eqnarray}
	\p_t\bm{v}+ (\bm{v}\cdot {\bm \nabla})\bm{v} 
    = -\bm\nabla p +\nu_o\Delta\bm{v}^*\,.
 \label{eq:Euler2}
\end{eqnarray}

Bulk hydrodynamic equations (\ref{eq:incompressibility}) and (\ref{eq:Euler2}) must be supplemented by boundary conditions. For a free surface we should use one kinematic and two dynamic boundary conditions
\begin{eqnarray}
	(\p_t \Gamma)_n &=& v_n\Big|_\Gamma\,,
 \label{eq:KBC} \\
 	T_{ij}n_j\Big|_\Gamma &=& 0\,,
 \label{eq:DBC}
\end{eqnarray}
where $\bm{n}$ is the unit vector normal to the free 1d surface $\Gamma=\p \mathcal{M}$ of the 2d fluid domain $\mathcal{M}$. The kinematic boundary condition (KBC), Eq. (\ref{eq:KBC}), states that the velocity of the free surface in its normal direction is equal to the normal component of the velocity flow taken at the surface. The set of two dynamical boundary conditions (DBC) given by (\ref{eq:DBC}) imposes that both components of stress force acting on the segment of the surface vanish. These conditions are appropriate for interfaces with vacuum or air, assuming that the latter cannot maintain non-vanishing forces on the surface of the fluid.

For a rather general class of fluid flows it is not possible to satisfy both boundary conditions (\ref{eq:DBC}) with the stress tensor (\ref{eq:Tij}) by smooth velocity configurations. A singular boundary layer is formed. One can see it, for example, in a linear approximation \cite{abanov2018odd} and the phenomenon is very similar to a formation of a boundary layer for fluid with infinitesimal shear viscosity \cite{lamb1932hydrodynamics}. A non-vanishing shear viscosity $\nu_e$ or finite compressibility, characterized by a finite sound velocity $v_s$ result in a finite thickness of the boundary layer proportional to $\sqrt{\nu_e}$ \cite{abanov2018odd} or to $1/v_s$ \footnote{Abanov, Can, Ganeshan, Monteiro, to be published.}, respectively. If one assumes that at least for finite times the boundary layer is stable and very thin, the motion of the fluid surface should be defined by effective boundary conditions imposed on the interior part of the fluid. Colloquially speaking, the latter boundary conditions can be obtained by ``integrating out'' boundary layer. As a result, instead of two independent DBC (\ref{eq:DBC}), one should consider a single effective normal dynamic boundary condition
\begin{equation}
	\tilde{p}\Big|_\Gamma \equiv p-\nu_o\omega\Big|_\Gamma = 2\nu_o \p_s v_n\,,
 \label{eq:DBCmod}
\end{equation}
where $\p_sv_n=-n_i^*\p_iv_n$ is the derivative of normal velocity along the boundary and we introduced a notation $\tilde p$ -- pressure modified by vorticity $\omega = \p_iv_i^*$. 

While the precise way in which the tangent stress part of DBC (\ref{eq:DBC}) is satisfied depends on the exact structure of the boundary layer, here we show that the effective normal stress boundary condition is universal and is given by (\ref{eq:DBCmod}). We obtain this universal statement by taking a variational principle for ideal incompressible fluid and by modifying the hydrodynamic action by the lowest in gradients boundary term which breaks parity but preserves other symmetries of the system. We show that this boundary term produces (\ref{eq:DBCmod}) justifying the expectation of universality.

Let us start by rewriting (\ref{eq:Euler2}) as
\begin{eqnarray}
	\p_t\bm{v}+ (\bm{v}\cdot {\bm \nabla})\bm{v} 
    = -\bm\nabla \tilde p 
 \label{eq:Euler3}	
\end{eqnarray}
using the incompressibility of the fluid (\ref{eq:incompressibility}). The equation (\ref{eq:Euler3}) is indistinguishable from the conventional Euler equation~\footnote{Remember that in incompressible fluids pressure $p$ is not a thermodynamic variable, but, instead, is fully determined by the flow $\bm{v}$. From this point of view $p\to\tilde p$ is just a change of notations.}. Therefore, we can start from the Luke's variational principle to produce the bulk hydro equations together with perfect fluid boundary conditions and look for boundary corrections to LVP to obtain the modified boundary conditions on the fluid which are in agreement with (\ref{eq:DBCmod}).

In contrast with \cite{abanov2018odd}, here we do not use any expansions in $\nu_e$ and our results do not rely on small surface angle approximations or on assumptions on the structure of the boundary layer.

\paragraph{\textbf{Luke's variational principle}.}

Let us start from the simplest case of the incompressible potential fluid flow, that is, $\bm v =\bm\nabla \theta$. Luke's variational principle is written in terms of the velocity potential $\theta$ as follows
\begin{equation}
	S_{\mathcal M} =-\int dt \int_{\mathcal M}d^2x\, \left(\p_t\theta+\frac{1}{2}(\p_i\theta)^2 \right) \,,
 \label{eq:LVP}
\end{equation}
where $\mathcal M$ is the 2D fluid domain with boundary. Variation over $\theta$ in the bulk gives $\Delta\theta=0$ -- the incompressibility condition. It is also straightforward to obtain (\ref{eq:Euler3}) as an identity if the modified pressure is identified as 
\begin{eqnarray}
	\tilde p = -\p_t\theta-\frac{1}{2}(\p_i\theta)^2\,.
 \label{eq:tildep}
\end{eqnarray}
Thus, the action (\ref{eq:LVP}) produces both bulk equations (\ref{eq:incompressibility}) and (\ref{eq:Euler3}) for a potential flow. The bulk vorticity of such flow vanishes identically $\omega =0$, implying $\tilde p = p$. Let us now keep track of boundary terms and assume that the bulk equation of motion $\Delta\theta=0$ is satisfied. Hence, varying (\ref{eq:LVP}) over the velocity potential $\theta$ and over shape of the fluid domain $\mathcal M$, we obtain that all the non-trivial dynamics resides on the fluid boundary and the action variation becomes (for details see Supplemental Material):
\begin{eqnarray}
	\delta S_{\mathcal M} &=&\int dt \int_{\Gamma}ds\,
    \left(\delta\theta\Big[(\p_t \Gamma)_n-\p_n\theta\Big]
    +(\delta\Gamma)_n \tilde{p} \right)\,.
 \label{eq:ds27}
\end{eqnarray}
Here $\Gamma=\p\mathcal M$ is the spatial boundary of the fluid domain and $s$ is the natural parameter along the boundary so that $dx^2+dy^2=ds^2$. The variation over the boundary values of the potential $\theta$, i.e., the first term in the integrand gives the KBC (\ref{eq:KBC}). The variation of the boundary, i.e., the second term in the integrand in (\ref{eq:ds27}), gives the vanishing pressure boundary condition $\tilde p|_\Gamma=0$ well known for ideal fluids. The latter is markedly different from the effective DBC (\ref{eq:DBCmod}) derived in \cite{abanov2018odd}. Therefore, while the variational principle (\ref{eq:LVP}) produces all equations and boundary conditions for ideal fluid it does not account for the contributions from odd viscosity.

\paragraph{\textbf{Boundary term}.}
The main result of this work is that in order to obtain the effective dynamical boundary condition (\ref{eq:DBCmod}), the following boundary term should be added to LVP: 
\begin{eqnarray}
	S_\Gamma &=&\nu_o\int dt \int_{\Gamma}ds\, (\p_t\Gamma)_n \alpha\,,
 \label{eq:SGamma}
\end{eqnarray}
where $s$ is the natural parameter along the boundary and $\alpha$ is the angle between the surface and some fixed direction.
\begin{figure}
\centering
\includegraphics[width=0.45\textwidth]{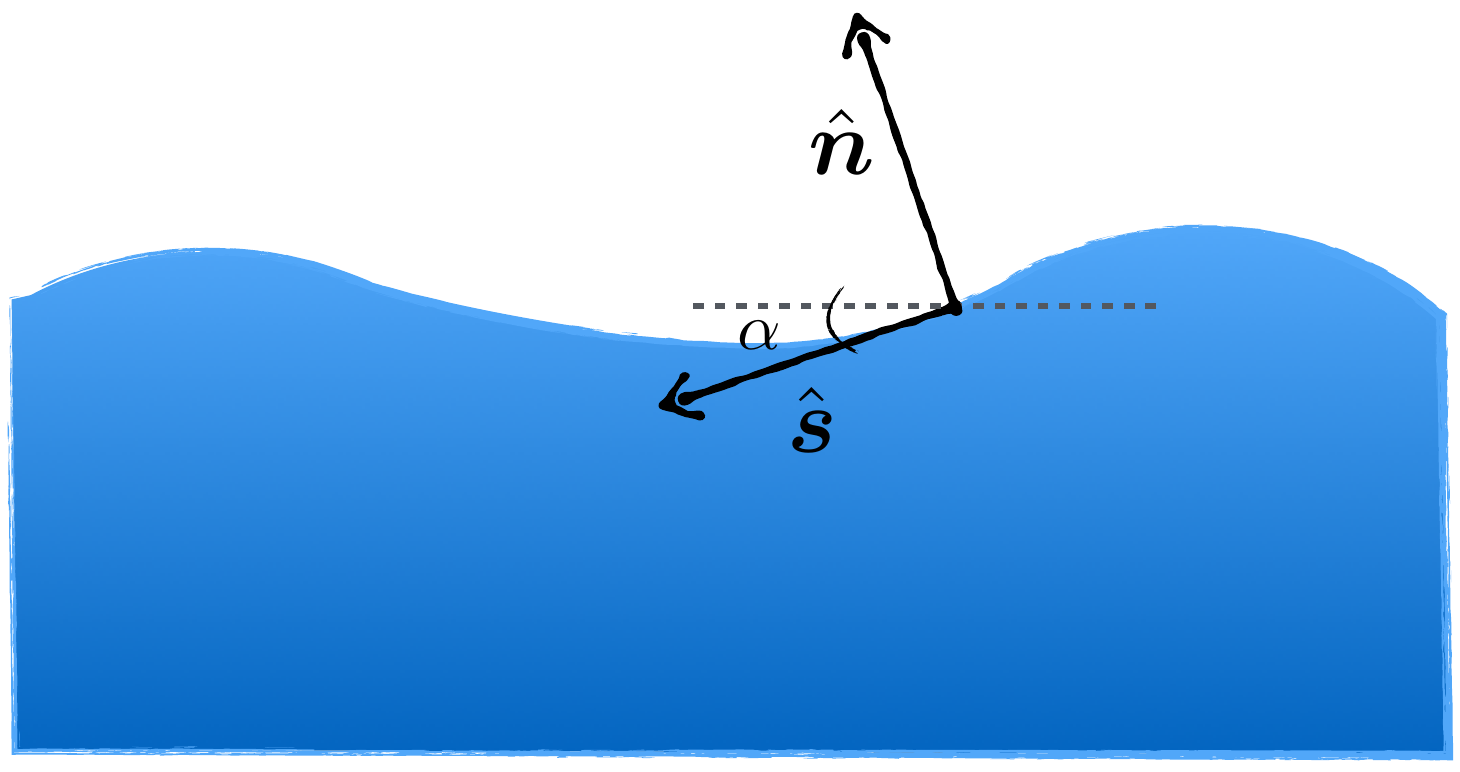}
\caption{\label{fig:axes}The choice of normal $\hat{n}$ and tangent $\hat{s}$ unit vectors used in the paper. In the case of half-plane geometry the angle $\alpha$ is measured from the horizontal axis as shown.}
\end{figure}
Two remarks are in order: (i) The term (\ref{eq:SGamma}) is constructed purely from the boundary geometry data and does not contain, e.g., velocity potential $\theta$. This means that the KBC (\ref{eq:KBC}) is not modified by this term. (ii)~Naively, (\ref{eq:SGamma}) contains the preferred direction -- the reference axis for $\alpha$. However, shifting $\alpha$ by constant does not change (\ref{eq:SGamma}) because $\int_\Gamma ds (\p_t\Gamma)_n=0$ due to the KBC and incompressibility of the fluid. Later on we will present a covariant way of writing (\ref{eq:SGamma}).

Let us first consider the example of a half-plane geometry when the fluid domain $\mathcal{M}$ is given by $y\leq h(x,t)$. We choose the reference direction to be the $x$-direction and write the angle $\alpha$ explicitly as $\alpha =\tan^{-1}h_x$. In this geometry $ds=\sqrt{1+h_x^2}\,dx$ and for normal velocity of the boundary we have
\begin{equation}
	(\p_t\Gamma)_n=\frac{h_t}{\sqrt{1+h_x^2}}=v_n\,, 
 \la{eq:KBC-h}
\end{equation}
so that Eq.~(\ref{eq:SGamma}) can be written as
\begin{eqnarray}
	S_\Gamma &=& \nu_o\int dt \int_{\mathbb R} dx\, h_t \alpha
    =-\nu_o\int dt \int_{\mathbb R} dx\, h \alpha_t,
 \label{eq:SGammah}
\end{eqnarray}
with $\alpha_t= \frac{h_{xt}}{1+h_x^2}$. Notice that in the last integral $h\,dx$ is the area element and $\alpha_t$ is the angular velocity of the surface element~\footnote{This observation makes the term $h \alpha_t$ very natural for representing odd viscosity. Remember that using (\ref{eq:tildep}) one can interpret (\ref{eq:LVP}) as a spacetime integral of modified pressure. The pressure is modified by $\nu_o\omega$ term and the vorticity $\omega$ is proportional to the angular velocity of local rotation of the fluid.}. Computing the variation of (\ref{eq:SGammah}) we obtain (for details see Supplemental Material):
\begin{align}
	     \delta S_\Gamma &= -2\nu_o\int dt \int_{\Gamma}ds\, 
    (\delta \Gamma)_n \,\p_s (\p_t \Gamma)_n\,,  
 \label{eq:SGh-var}
\end{align}
where $(\p_t \Gamma)_n$ is given by (\ref{eq:KBC-h}) and $(\delta\Gamma)_n = \delta h/\sqrt{1+h_x^2}$. 

It is easy to see that the variation $\delta(S_{\mathcal M}+S_{\Gamma})$ over $(\delta \Gamma)_n$ given by (\ref{eq:ds27}) and (\ref{eq:SGh-var}) gives the modified boundary condition (\ref{eq:DBCmod}).

The same analysis can be repeated for the geometry of a disk, i.e., simply connected droplet producing again (\ref{eq:SGh-var}) \footnote{Assuming that the boundary action (\ref{eq:SGamma}) is well defined (see the following sections) there is, actually, no need to repeat the calculations. Any surface can be locally parameterized as $y=h(x,t)$ and all variational calculations we perform are local. Nevertheless, we give explicit formulas for the droplet geometry in Supplemental Material for future references.}. Therefore, the variational principle with the action  
\begin{eqnarray}
	S = S_{\mathcal M}+S_\Gamma
 \label{eq:fullaction}
\end{eqnarray}
defined in (\ref{eq:LVP}) and (\ref{eq:SGamma}) produces incompressibility condition and the Euler equation (\ref{eq:Euler3}) with kinematic (\ref{eq:KBC}) and effective dynamic (\ref{eq:DBCmod}) boundary conditions. Explicitly, the full set of equations can be written as
\begin{align}
	\Delta \theta &= 0 \,, \quad & x\in \mathcal{M}\,,
 \label{eq:Laplace}\\
 	\p_n\theta &= (\p_t\Gamma)_n \,, \quad & x\in \Gamma\,,
 \label{eq:pntheta}\\
 	\p_t\theta &+\frac{1}{2}(\p_i\theta)^2 = -2\nu_o \p_s (\p_t\Gamma)_n\,, \quad & x\in \Gamma\,.
 \label{eq:pttheta}
\end{align}

The obtained hydrodynamics describes incompressible potential flows of the fluid with odd viscosity. This is the main result of this work. We will remove the requirement of potentiality of the flow later in this paper.   

\paragraph{\textbf{Effective contour dynamics}.} In the case of an irrotational bulk flow, the full dynamics is completely determined by the boundary motion. One can express equations (\ref{eq:pntheta},\ref{eq:pttheta}) purely in terms of boundary fields using (\ref{eq:Laplace}). To do that we introduce the boundary field $\tilde\theta=\theta|_\Gamma$ or explicitly $\tilde\theta(s,t)=\theta(x(s,t),y(s,t),t)$ with boundary $\Gamma$ given parametrically by functions of the natural parameter $s$ along the boundary. We use the identity
\begin{eqnarray}
	\p_t\tilde\theta = \p_t\theta|_\Gamma+(\p_n\theta)(\p_t\Gamma)_n
\end{eqnarray}
in (\ref{eq:pttheta}) together with (\ref{eq:pntheta}) and obtain
\begin{eqnarray}
	\p_t\tilde\theta +\frac{1}{2}(\p_s\tilde\theta)^2-\frac{1}{2}{(\p_t\Gamma)_n}^2 
    &=& -2\nu_o \p_s (\p_t\Gamma)_n\,.
 \label{eq:pttildetheta}
\end{eqnarray}
The equation (\ref{eq:pntheta}) can also be expressed in terms of boundary fields using (\ref{eq:Laplace}). It has a form
\begin{eqnarray}
	(\p_t\Gamma)_n &=& \widehat{DN}\tilde\theta\,,
 \label{eq:ptDN}
\end{eqnarray}
where $\widehat{DN}$ is a Dirichlet to Neumann operator which depends on the shape of the domain and can be expressed in terms of the Dirichlet Green function of Laplace operator\footnote{Alternatively one can invert the equation (\ref{eq:ptDN}) and write it as $\theta =\hat{S}(\p_t\Gamma)_n$ using the Neumann operator (see Supplemental Material).} (see Supplemental Material) as:
\begin{eqnarray}
	\widehat{DN}\,\tilde\theta(s) = \int_\Gamma ds'\, \big[\p_n\p_{n'} G(x,x')\big] \tilde\theta(s') \,. \la{DNdef}
\end{eqnarray}

For the case of the domain given by $y\leq h(x,t)$ one can find $\widehat{DN}$ as an expansion in $h$ and obtain \cite{abanov2018odd}
\begin{eqnarray}
	\widehat{DN}\theta = -\tilde{\theta}^H_x-\Big[h\tilde\theta_x+(h\tilde\theta_x^H)^H\Big]_x+\ldots,
 \label{eq:DNh}
\end{eqnarray}
where the Hilbert transform is defined as $f^H(x)=\dashint \frac{dx'}{\pi} \frac{f(x')}{x'-x}$.

The equations (\ref{eq:pttildetheta},\ref{eq:ptDN}) give a closed system of boundary dynamics of a droplet in terms of purely boundary fields. These are exact equations given by the action (\ref{eq:fullaction}). The approximate versions of these equations using (\ref{eq:DNh}) can be found in Ref.~\cite{abanov2018odd}.

It is even easier to derive the effective one-dimensional action corresponding to equations (\ref{eq:pttildetheta},\ref{eq:ptDN}). We integrate (\ref{eq:fullaction}) by parts and use the bulk incompressibility of the fluid $\Delta\theta=0$ to obtain (for details see Supplemental Material):
\begin{align}
	S_{1D} &=\int dt \left[\int_{\Gamma}ds\,
    (\p_t \Gamma)_n\left(\tilde\theta+\nu_o\alpha\right)- H\right]\,,
 \la{eq:1Daction} \\
 	H &=\frac{1}{2} \int_\Gamma ds\,\big(\theta\,\p_n\theta\big)_\Gamma
    =\frac{1}{2} \int_\Gamma ds\,\tilde\theta\,\widehat{DN}\tilde\theta \,.
 \la{eq:ham1}
\end{align}
The Hamiltonian (\ref{eq:ham1}) is nothing but the total kinetic energy of the fluid given by the second term of (\ref{eq:LVP}).
The variations of (\ref{eq:1Daction}) with respect to $\tilde\theta$ and displacements of the boundary produce equations of motion (\ref{eq:pttildetheta},\ref{eq:ptDN}). These variations can be computed using  the Hadamard's variational formula, defined in \cite{Warschawski-Hadamard, PEETRE-Hadamard}. For further details, see the Supplemental Material.


\paragraph{\textbf{Hamiltonian structure of contour dynamics}.} 
Instead of studying the boundary dynamics for a general fluid domain $\mathcal M$, let us focus here on the particular case when $\mathcal M$ is given by $y\leq h(x,t)$. Then, the action (\ref{eq:1Daction}) can be rewritten as
\begin{equation}
	S_{1D} =\int dt\left[\int_{\mathbb R} dx\,h_t (\tilde\theta+\nu_o\alpha)-H\right]\,,
 \la{eq:1Daction25}
\end{equation}
where the Hamiltonian is given by (\ref{eq:ham1}), with $ds=\sqrt{1+h_x^2}\,dx$. Let us turn our attention to the first term of (\ref{eq:1Daction25}). We immediately see that $h$ and $\tilde\theta-\nu_o\alpha$ are canonically conjugated variables so that Poisson brackets become \footnote{We use abbreviated notations $h=h(x)$, $h'=h(x')$ etc. For details of the derivation see Supplemental Material.}
\begin{align}
	\left\{h,h'\right\} &= 0\,,
 \label{eq:PBhh} \\
  	\left\{\tilde\theta,h'\right\} &= \delta(x-x')\,,
 \label{eq:PBthetah} \\
	\left\{\tilde\theta,\tilde\theta'\right\} 
    &= \nu_o\left(\frac{1}{1+h_x^2}+\frac{1}{1+{h_{x}'}^2}\right) \p_x\delta(x-x')\,.
 \label{eq:PBthetatheta}
\end{align}
Note that that the Poisson structure reduces to the well known Zakharov's Poisson structure \cite{zakharov1968stability} when $\nu_o=0$. In the limit of small slopes $h_x\ll 1$ the bracket (\ref{eq:PBthetatheta}) was obtained in \cite{abanov2018odd}. However, we emphasize here that the Poisson structure (\ref{eq:PBhh},\ref{eq:PBthetah},\ref{eq:PBthetatheta}) is an exact consequence of the variational principle (\ref{eq:fullaction},\ref{eq:LVP},\ref{eq:SGammah}) without any additional approximations.

\paragraph{\textbf{Bulk vorticity}.} 
It is straightforward to generalize the variational principle for the case when vorticity might be present in the bulk. We simply replace the bulk action (\ref{eq:LVP}) by
\bea
	S_{\mathcal M} &=& -\int dt\,\int_{\mathcal M} d^2x\, \Big[\xi_0
	+  \frac{\xi_i \xi^i}{2} \Big]\,
 \la{eq:action121}
\eea
without changing the boundary action (\ref{eq:SGamma}). Here, 
\bea
	\xi_\mu \equiv \p_\mu\theta 
    +\Phi \p_\mu \Psi \,,
 \label{cleb121}
\eea
$\Phi,\Psi$ are additional Clebsch parameters \cite{mobbs1982variational} and $\theta,\Phi,\Psi$ are considered as independent variational fields. It is easy to see that the variational principle (\ref{eq:fullaction}) with (\ref{eq:action121},\ref{eq:SGamma}) produces correct bulk equations for $v_i=\xi_i$ and $\tilde{p}=-\left(\xi_0+\tfrac{1}{2}\xi_i^2\right)$, as well as the kinematic boundary condition (\ref{eq:KBC}) and the effective dynamic boundary condition (\ref{eq:DBCmod}). The obtained dynamics is again Hamiltonian with Poisson structure given by the deformation of the brackets of Ref.~\cite{lewis1986hamiltonian} by boundary terms proportional to $\nu_o$.

The general dynamics (\ref{eq:fullaction},\ref{eq:action121},\ref{eq:SGamma}) cannot be reduced to the dynamics of the surface as it includes non-trivial evolution of bulk vorticity. There are, however, few cases when nontrivial contour dynamics can be obtained. These cases include dynamics of potential flows in multiply connected domains or domains with free boundaries and patches of constant vorticity. In both these cases the effective dynamics is essentially the one of coupled contours - domain boundaries and boundaries of vorticity patches. We will postpone the studies of these cases for future. Another interesting case is the interaction between point vortices in the bulk of the fluid with the free surface.

\paragraph{\textbf{Boundary term and geometry}.} The boundary term (\ref{eq:SGamma}) involves some arbitrariness in choosing a reference direction. In this paragraph, we aim to give a more covariant way of this form and to provide a geometrical picture associated with this boundary action. For that, it is convenient to express $S_\Gamma$ in terms of differential forms. Since $\bm{\hat n}=(\sin\alpha,-\cos\alpha)$, we can associate the derivatives of the angle $\alpha$ to the boundary extrinsic curvature one-form $K=K_\mu dx^\mu$ (for details, vide \cite{gromov2016boundary})
\begin{equation}
	K_\mu=n_i\p_\mu s_i=n_i\p_\mu n^*_i=\p_\mu\alpha \,.
\end{equation}
Integrating by parts, we can rewrite (\ref{eq:SGamma}) as
\begin{equation}
	S_\Gamma=-\nu_o\int_{\mathbb R\times \Gamma}A\wedge K \,, 
 \la{SGamma-forms}
\end{equation}
where $A$ is a one-form whose exterior derivative is the plane volume-form, that is, $dA=dx\wedge dy$. There is an ambiguity in the definition of $A$, since $A'=A+d\Lambda$ gives us $dA'=dA$. However, this gauge freedom does not affect the boundary action (\ref{SGamma-forms}). \footnote{It is assumed here that $\Lambda$ is single-valued. To allow for large gauge transformations one should include additional bulk terms involving spin connection.  \cite{gromov2016boundary}}

As an example let us consider $A=-ydx$ for $\mathcal{M}$ given by $y\leq h(x,t)$. Then, Eq. (\ref{SGamma-forms}) reproduces (\ref{eq:SGammah}).

For the droplet case, $\mathcal{M}$ is defined (in polar coordinates) by $r\leq R(\varphi,t)$. If we take $A=\frac{1}{2}r^2\,d\varphi$, we then obtain:
\begin{eqnarray}
	S_{\Gamma} =- \frac{\nu_o}{2}\int_{\mathbb R\times\Gamma}R^2 \alpha_t\, dt\wedge d\varphi \,.
 \label{eq:Somega20}
\end{eqnarray}

\paragraph{\textbf{Conclusions}.} We presented a variational principle which accounts for odd viscosity effects in incompressible fluid dynamics. The boundary part of the proposed action is purely geometrical and fully determined by the symmetries of the system. Therefore, we expect the boundary condition (\ref{eq:DBCmod}) to be universal and independent on the exact structure of the boundary layer, given this boundary layer to be sufficiently thin. In particular, Eq. (\ref{eq:DBCmod}) reproduces the approximate equations obtained in Ref.~\cite{abanov2018odd}, which were derived in the limit of very small, but nonvanishing shear viscosity. We also expect the same boundary conditions assuming the boundary layer structure to be determined by a finite compressibility of the fluid. If the fluid is compressible, the odd viscosity affects the flow of the fluid in the bulk as well. While it is relatively straightforward to construct a variational principle for the compressible fluid its connection to the incompressible limit is subtle and will be discussed elsewhere. 

The variational principle (\ref{eq:fullaction},\ref{eq:action121},\ref{eq:SGamma}) gives hydrodynamic equations for an incompressible fluid with odd viscosity under the assumption that the tangent stress free surface boundary conditions can be satisfied by a thin boundary layer. This is not the case for all fluid flows. For example, in the geometry of an expanding air bubble exact solutions show strong dependence of the bulk flow on shear viscosity \cite{ganeshan2017odd}. Also, even if the assumption of a thin boundary layer is satisfied initially it might break at finite time \cite{abanov2018odd}. The applicability of the thin boundary layer assumption is beyond of the scope of this letter. 

 In the irrotational case, the degrees of freedom reside on the boundary and the effective dynamics is one-dimensional and Hamiltonian, albeit non-local. The derived Hamiltonian structure modifies the well known Hamiltonian structure of incompressible ideal fluids \cite{zakharov1968stability}.






\paragraph{\textbf{Acknowledgements}.}
We are grateful to Sriram Ganeshan and Paul Wiegmann 
for many fruitful discussions and suggestions related to this project. AGA’s work was supported by grant NSF DMR-1606591. GMM thanks Funda\c c\~ao de Amparo \`a Pesquisa do Estado de S\~ao Paulo (FAPESP) for financial support under grant 2016/13517-0.

\bibliographystyle{my-refs}
\bibliography{hydro-review-bibliography}

\mbox{}
\vspace{5cm}

\newpage
\newpage
\appendix

\section{Supplemental Material}
 \label{app:td}

\subsection{Computing variations}

The action (\ref{eq:LVP}) depends on two variables: velocity potential $\theta$ and the shape of the domain $\Gamma=\partial{\mathcal M}$. Equations of motions are obtained by varying (\ref{eq:LVP}) in terms of both variables, keeping track of boundary terms. Thus, we have
\begin{eqnarray}
	\delta S_{\mathcal M} &=& -\int dt \int_{\mathcal M}d^2x\, \Big(\p_t(\delta\theta)
    +(\p_i\theta)\p_i(\delta\theta) \Big)  
 \nonumber \\
 	&-& \int dt \int_{\Gamma}ds\, (\delta \Gamma)_n\Big(\p_t\theta+\frac{1}{2}(\p_i\theta)^2 \Big)
 \nonumber \\
 	&=& -\int dt \int_{\mathcal M}d^2x\, \Big(\p_t(\delta\theta)
    +\p_i(\delta\theta \p_i\theta)-\delta\theta \Delta\theta \Big)
 \nonumber \\
 	&+& \int dt \int_{\Gamma}ds\, (\delta \Gamma)_n \tilde{p} \,,
 \label{eq:LVPvar1}
\end{eqnarray}
where we used Eq. (\ref{eq:tildep}). Integrating full derivatives in (\ref{eq:LVPvar1}) and using the Leibniz integral rule, we obtain
\begin{eqnarray}
	\delta S_{\mathcal M} 
 	&=& \int dt \int_{\mathcal M}d^2x\,\delta\theta \Delta\theta 
 \label{eq:LVPvar10} \\
    &+&\int dt \int_{\Gamma}ds\,
    \left(\delta\theta\Big[(\p_t \Gamma)_n-\p_n\theta\Big]
    +(\delta\Gamma)_n \tilde{p} \right) \,.
 \nonumber
\end{eqnarray}
The first line gives the bulk incompressibility condition $\Delta \theta =0$ and the second line reproduces (\ref{eq:ds27}). 

Let us now compute the variation of (\ref{eq:SGammah}). For that, we have
\begin{align}
	\delta S_\Gamma &= -\nu_o \int dt \int_{\mathbb R} dx\, (\delta h \alpha_t -h_t\delta\alpha)\,,
  \nonumber \\
    &= -\nu_o \int dt \int_{\mathbb R} dx\, \left(\delta h \frac{h_{xt}}{1+h_x^2} 
    -h_t\frac{\delta h_x}{1+h_x^2}\right)\,,
 \nonumber \\
    &= -2\nu_o \int dt \int_{\mathbb R} dx\, \delta h\left[ \frac{h_{xt}}{1+h_x^2} 
    -\frac{h_t h_x h_{xx}}{(1+h_x^2)^2}\right] \,.
 \label{eq:SGammahvar1}
\end{align}
Straightforward manipulations show that (\ref{eq:SGammahvar1}) can be written as
\begin{align}
	\delta S_\Gamma &=-2\nu_o\int dt \int_{\mathbb R} dx\, 
     \frac{\delta h}{\sqrt{1+h_x^2}}\,\p_x\left(\frac{h_t}{\sqrt{1+h_x^2}}\right)\,,
 \label{eq:deltaSGammahvar}
\end{align} 
which is equivalent to (\ref{eq:SGh-var}) of the main text. In principle, we can always write the boundary equation as $y-h(x,t)=0$ locally, therefore, this result is general. To see that, let us consider a droplet $\mathcal{M}$ --which is topologically a disk. In polar coordinates, $\mathcal{M}$ is defined by $r\leq R(\varphi,t)$. To guarantee that $\alpha$ is sigle valued, let us choose the reference direction to be the radial one, such that, $\alpha =\tan^{-1}(R/R_\varphi)$. The kinematic boundary condition becomes
\begin{equation}
	(\p_t\Gamma)_n=\frac{R R_t}{\sqrt{R^2+R_\varphi^2}}=v_n\,, 
 \la{eq:KBC-R}
\end{equation}
and the boundary action (\ref{eq:SGamma}) can be re-expressed as
\begin{equation}
	S_\Gamma = -\nu_o\int dt \int\limits_0^{2\pi} d\varphi\, RR_t \alpha
    =\nu_o\int dt \int\limits_0^{2\pi} d\varphi\, \frac{R^2}{2} \alpha_t \,.
 \label{eq:SGammaR}
\end{equation}
Varying (\ref{eq:SGammaR}) over $R$ we have
\begin{align}
	\delta S_\Gamma &= \nu_o\int dt \int\limits_0^{2\pi} d\varphi\, 
    (R\delta R \alpha_t -\delta\alpha RR_t)\,,
 \nonumber \\
   & = \nu_o\int dt \int\limits_0^{2\pi} d\varphi\, R^2\,
   \frac{\delta R_\varphi R_t-\delta RR_{\varphi t}}{R^2+R_\varphi^2}\,.
 \nonumber
\end{align}
    Integrating this equation by parts, we obtain
    \begin{align}
   & \delta S_\Gamma =  \label{eq:SGammaRvar1}
   \\
  & 2\nu_o\int dt \int\limits_0^{2\pi} d\varphi\, \delta R\left[\frac{ R_{ t}R_\varphi(R^2 R_{\varphi\varphi}-RR_\varphi^2)}{(R^2+R_\varphi^2)^2}-\frac{ R^2R_{\varphi t}}{R^2+R_\varphi^2}\right],\nonumber
\end{align}
and after some straightforward, but rather tedious, algebra, Eq. (\ref{eq:SGammaRvar1}) can be expressed as:
\begin{align}
    \delta S_\Gamma &=-2\nu_o\int dt \int\limits_0^{2\pi} d\varphi\, 
     \frac{R \delta R}{\sqrt{R^2+R_\varphi^2}}\,\p_\varphi\left(\frac{R R_t}{\sqrt{R^2+R_\varphi^2}}\right),
 \label{eq:SGR-var}
\end{align}
which can be brought to the form (\ref{eq:SGh-var}) using (\ref{eq:KBC-R}).

\subsection{Potential problem}

The velocity potential is a harmonic function and can be expressed solely through its boundary value $\theta|_{\Gamma}$. For that, we must solve the Laplace equation $\Delta\theta=0$ on $\mathcal{M}$ with Dirichlet boundary condition $\theta|_{\Gamma}=\tilde\theta(s,t)$. This solution is formally given in terms of the Laplace Green's function, that is,
\begin{align}
\Delta_{\bm{x}}G(\bm x, \bm x')=\delta(\bm x-\bm x'),& \qquad \bm x\in \mathcal M,
\\
G(\bm x, \bm x')=0,& \qquad \bm x\in\Gamma.
\end{align}
Thus,
\begin{equation}
\theta(\bm x, t)=\int_\Gamma ds'\, \p_{n'}G(\bm x, \bm x'(s',t))\,\tilde\theta(s',t)
\end{equation}
and, consequently,
\begin{equation}
\p_n\theta|_\Gamma=\int_\Gamma ds'\, \p_n\p_{n'}G(\bm x(s,t), \bm x'(s',t))\,\tilde\theta(s',t).
\end{equation}

Here it is convenient to introduce the Dirichlet to Newman operator, defined in (\ref{DNdef}), that is,
\begin{equation}
\p_n\theta|_\Gamma\equiv\widehat{DN}\,\tilde\theta(s,t).
\end{equation}

In the one-dimensional effective dynamics, the Hamiltonian assumes an exact, yet non-local form:
\begin{equation}
H=\frac{1}{2}\int_\Gamma ds\,\tilde\theta(s,t)\,\widehat{DN}\, \tilde\theta(s,t). \la{eq:ham3}
\end{equation}
This can be obtained from the action term:
\begin{equation*}
\int_{\mathcal M} d^2x\,(\partial_i\theta)^2=\int_\Gamma ds \,(\theta\,\p_n\theta)_\Gamma-\int_{\mathcal M} d^2x\;\theta\,\Delta\theta\,,
\end{equation*}
after imposing the incompressibility condition 
\[\p_iv_i=\Delta\theta=0.\]
Variation of Hamiltonian (\ref{eq:ham3}) in terms of the boundary field $\tilde \theta$ is trivial and is given by:
\begin{equation}
\frac{\delta H}{\delta\tilde\theta(s,t)}=\widehat{DN}\, \tilde\theta(s,t)=v_n\,. 
\end{equation}

To obtain the Hamiltonian variation with respect to the boundary shape, it is necessary to know how the Green's function varies with respect to the boundary change. This variation is called Hadamard formula \cite{Warschawski-Hadamard, PEETRE-Hadamard} and can be written as:
\be
\delta G(\bm x,\bm y)=\int_\Gamma ds'\p_{n'}G(\bm x,\bm x')\p_{n'}G(\bm y,\bm x')(\delta\Gamma)_{n'}. \la{Hadamard}
\ee



In the following appendix, we will use this formula to obtain the equations of motion for the action (\ref{eq:1Daction25}).

\subsection{1D action, Hamiltonian, and Poisson brackets, Equations of motion}

For the domain $\mathcal M$ given by $y\leq h(t,x)$, Eq. (\ref{eq:1Daction25}) can be rewritten as:
\be
S_{1D}=\int dt\left[\int dx\,\pi(x,t) h_t(x,t)-H\right].
\ee

Choosing the reference direction to be the $x$ axis, the canonical momentum is given explicitly by:
\be
\pi(x,t)=\tilde\theta(x,t)+\nu_o\tan^{-1}h_x(x,t) \la{pi}
\ee
and, the Poisson algebra for such system becomes trivial:
\begin{align}
\{h(x,t),h(x',t)\}&=0, \la{hh}
\\
\{\pi(x,t),\pi(x',t)\}&=0, \la{pp}
\\
\{\pi(x,t),h(x',t)\}&=\delta(x-x'). \la{ph}
\end{align}

In order to shorten up the notation, let us denote $h(x',t)$ and $\tilde \theta(x',t)$ as $h'$ and $\tilde\theta'$ respectively. Equations (\ref{hh} - \ref{ph}) together with (\ref{pi}) imply that:
\begin{align}
\{\pi,h'\}&=\{\tilde\theta,h'\}=\delta(x-x'),
\\
\{\tilde\theta,\tilde\theta'\}&=-\nu_o\{\tilde\theta,\tan^{-1}h_{x}'\}-\nu_o\{\tan^{-1}h_x,\tilde\theta'\}\nonumber
\\
\{\tilde\theta,\tilde\theta'\}&= \nu_o\left(\frac{1}{1+h_x^2}+\frac{1}{1+{h_{x}'}^2}\right) \p_x\delta(x-x').
\end{align}

Thus, Hamilton equations for the system are given by:
\begin{align}
h_t&=\{H,h\}=\frac{\delta H}{\delta\tilde\theta}, \la{eom-h}
\\
\tilde\theta_t&=\{H,\tilde\theta\}, \nonumber
\\
\tilde\theta_t&=-\frac{\delta H}{\delta h}-\frac{2\nu_o}{\sqrt{1+h_x^2}}\p_x\left(\frac{1}{\sqrt{1+h_x^2}}\frac{\delta H}{\delta\tilde\theta}\right). \la{eom-theta}
\end{align}

Using that $ds=\sqrt{1+h_x^2}\, dx$, we find that:
\be
H=\frac{1}{2}\int dx\,\sqrt{1+h_x^2}\,\tilde\theta(x)\widehat{DN}\tilde\theta(x),
\ee
and the equation of motion (\ref{eom-h}) becomes Eq. (\ref{eq:KBC-h}):
\be
\frac{h_t}{\sqrt{1+h_x^2}}=\widehat{DN}\tilde\theta=v_n.
\ee


\subsection{Wen-Zee term with boundary correction}


Let us consider the following geometric action as proposed in \cite{gromov2016boundary}:
\begin{eqnarray}
	S_{\omega} = \nu_o\int_{\Sigma} A \wedge d\omega +\nu_o\int_{\p\Sigma}A\wedge K \,, \la{WZ}
\end{eqnarray}
where $\Sigma$ is the Newton-Cartan spacetime domain. This action was originally regarded to account for boundary effects in quantum Hall systems. The first term is the so-called Wen-Zee term, where $\omega$ is the spin connection and $A$ is the gauge field. In the presence of boundary, Wen-Zee term is not invariant under gauge transformations $A\to A+d\Lambda$, however the full action (\ref{WZ}) is. Moreover, it is also invariant with respect to frame rotations. Integrating by parts we obtain
\begin{eqnarray}
	S_{\omega} = \nu_o\int_{\Sigma} \omega\wedge dA +\nu_o\int_{\p\Sigma}A\wedge (K-\omega) \,.
 \label{eq:Somega}
\end{eqnarray}

In the case of Hall effect in a flat background, $dA$ is simply the magnetic field multiplied by the area element $dx\wedge dy$. Setting the magnetic field equals to 1, the bulk term becomes:
\be
\int_{\Sigma} \omega\wedge dA =\nu_o\int_{\Sigma}\omega_0 \, dt\wedge dx\wedge dy.
\ee

It was shown in \cite{gromov2016boundary} that the spin connetion projected on the boundary satisfies:
\be
\omega|_\Gamma=K+d\alpha.
\ee

Then (\ref{eq:Somega}) becomes 
\begin{eqnarray}
	S_{\omega} = \nu_o\int_{\mathcal M} \omega_0\,dV -\nu_o\int_{\partial\mathcal{M}}A\wedge d\alpha \,,
 \label{eq:Somega2}
\end{eqnarray}
where $\alpha$ is the angle of tangent to the boundary to some fixed direction. The second term (\ref{eq:Somega2}) can be easily recognized as (\ref{eq:SGamma}) while the first one gives the correction to spin density in the bulk. In flat background, we can always set $\omega_0=0$.

As an example let us consider $A=-ydx$ for $\mathcal{M}$ given by $y\leq h(x,t)$. The second term of (\ref{eq:Somega2}) becomes
\begin{eqnarray}
	S_{\omega} = \nu_o\int_{\partial\mathcal{M}}h \alpha_t\, dt\wedge dx \,.
 \label{eq:Somega100}
\end{eqnarray}

We could have in principle chosen any other one-form $A'$, such that $dA= dx\wedge dy$, and the action would have been unchanged, due to its gauge invariance. For example, using $A=\tfrac{1}{2}(x\, dy- y\,dx)$, we have
\[
A|_{\partial\Sigma}=\frac{1}{2}[xh_t dt+(xh_x-h)dx],
\]
and 
\begin{align}
A|_{\p\Sigma}\wedge d\alpha&=\frac{1}{2}[xh_t dt+(xh_x-h)dx]\wedge[\p_t\alpha\, dt+\p_x\alpha\, dx]\nonumber
\\
&=\left[h\p_t\alpha-\frac{1}{2}\p_x\left(x h\p_t\alpha\right)+\frac{1}{2}\p_t\left(x h \p_x\alpha\right)\right]dt\wedge dx
\end{align}

In general, $A$ is defined up to and exact form $d\Lambda$, therefore:
\begin{equation}
A'\wedge d\alpha=(A+d\Lambda)\wedge d\alpha=A\wedge d\alpha+d(\Lambda d\alpha).
\end{equation}

For the droplet case $r\leq R(\phi,t)$, it is convenient to take $A=\frac{1}{2}r^2\,d\phi$. We have then from (\ref{eq:Somega2})
\begin{eqnarray}
	S_{\omega} = \frac{\nu_o}{2}\int_{\partial\Sigma}R^2 \alpha_t\, d\phi\wedge dt \,.
 \label{eq:Somega200}
\end{eqnarray}

\end{document}